\documentclass{article}

\usepackage{spconf}
\usepackage{amsmath}
\usepackage{graphicx}
\usepackage{tabularx}
\usepackage{booktabs}
\usepackage{multirow}
\usepackage{cite}
\usepackage{hyperref}
\usepackage{xcolor}
\usepackage{enumitem}
\usepackage{etoolbox}
\usepackage{microtype}
\AtBeginDocument{\microtypesetup{activate=false}}

\hypersetup{
    hidelinks
}

\title{HiThink Turn: An Intent-Aware Turn-Taking Control Module
for Full-Duplex Dialogue}

\name{%
\begin{tabular}{@{}c@{}}
Feiyang Chen$^{*}$,
Wenhan Yang$^{*}$,
Bohan Wang,
Xinjian Gao, \\
Rongjunchen Zhang,
Jun Wang,
and Xinhui Hu$^{\dagger}$
\end{tabular}%
\thanks{$^{*}$These authors contributed equally.
\quad $^{\dagger}$Corresponding author.}
}

\address{
HiThink Research, Hangzhou, China \\
\texttt{\{chenfeiyang2,yangwenhan,huxinhui\}@myhexin.com}
}

\begin{document}


\maketitle

\begingroup
\setlength{\parskip}{0pt}
\microtypesetup{activate=true}
\begin{abstract}
Full-duplex dialogue requires timely yet selective interruption handling, which end-of-turn prediction alone cannot achieve: complete utterances may need no response, while unfinished requests may warrant interruption. To address this challenge, we propose HiThink Turn, an intent-aware streaming turn-state predictor that separates response intent from semantic completeness and conditions decisions on system playback state. A key contribution is minimal intent-sufficient prefix supervision, constructed through LLM judgments and speech alignment, while training on audio truncated at chunk boundaries improves robustness to partial speech. These components support streaming inference with 240-ms audio chunks, enabling low-latency, accurate full-duplex turn control. Experiments show that HiThink Turn leads the compared methods in Easy Turn macro accuracy, Full-Duplex-Bench average interaction rate score (0.933), and non-target-speech average playback resume rate (0.735). Additionally, intent-prefix triggering raises interruption success from 89\% to 98\% and reduces mean stop latency by 60.9\%.
\end{abstract}


\section{Introduction}
\label{sec:intro}

Conventional spoken dialogue systems are half-duplex and turn-based, limiting their ability to handle interruptions and overlapping speech \cite{lu2026survey}. Full-duplex spoken dialogue systems (FD-SDSs) enable simultaneous listening and speaking through cascaded \cite{chen2025fireredchat,wang2025freezeomni} or end-to-end architectures \cite{defossez2024moshi,chen2025minmo}. Cascaded systems remain attractive for their modularity, reuse of mature components, and preservation of text-based LLM reasoning \cite{yang2026duplexcascade,soulxduplug2026,fu2026x2turn}. Their turn-state predictors determine whether to listen, respond, provide a backchannel, or stop playback, requiring both accuracy and low latency.

Early turn-state predictors primarily determine whether the current utterance has reached a suitable completion point. For instance, Smart Turn \cite{pipecatai2025smartturn} and TEN Turn Detection~\cite{tenteam2025turndetection} infer utterance completion from speech and ASR text, respectively. FlexDuo \cite{liao2025flexduo} and Easy Turn \cite{li2026easy} extend state modeling to irrelevant input, incomplete speech, and backchannels. Recent studies suggest that completion alone is insufficient for reliable interaction control: a semantically complete statement or monologue does not necessarily warrant a response \cite{li2026decoupling}. TurnFSM \cite{lin2026turnfsm} and Joy-Duplex \cite{bai2026joyaitalker} therefore further separate utterance completion from subsequent acceptance or rejection decisions, enabling finer-grained dialogue control.

Deciding whether to respond, however, does not by itself specify when sufficient evidence becomes available. During system playback, an unfinished request may already justify interruption, making utterance completion an unnecessarily late trigger. Streaming prediction from fixed-duration windows~\cite{wu2025phoenixvad}, audio chunks~\cite{soulxduplug2026}, or intermediate ASR outputs~\cite{yang2026duplexcascade} enables earlier decisions, but partial inputs can also lead to premature interruptions. The challenge is therefore to distinguish prefixes that already convey sufficient response intent from those requiring further listening. This motivates explicit intent-prefix supervision together with training on audio truncated at streaming boundaries.

To address these limitations, we propose HiThink Turn, an intent-aware
streaming turn-state predictor for cascaded FD-SDSs. Our main
contributions are:
\begin{itemize}[leftmargin=*, nosep, topsep=3pt]
    \item A state formulation that separates response intent from
    semantic completeness and conditions decisions on system playback state,
    distinguishing when to respond, preserve playback, or interrupt.

    \item A streaming supervision pipeline that labels the minimal intent-sufficient prefixes using LLM judgments and speech alignment,
    and combines boundary-aware truncation with non-target speech
    augmentation for 240-ms chunk-wise inference.
\end{itemize}

HiThink Turn is implemented by fine-tuning
Qwen3-ASR-1.7B~\cite{shi2026qwen3asr} using
LoRA~\cite{hu2022lora}. Results on Easy Turn and
Full-Duplex-Bench show accurate prediction of utterance
completeness, effective turn control, and robust rejection
of non-target speech. Triggering on intent prefixes improves
response success and reduces interruption latency compared
with triggering on utterance completion.

\par
\endgroup

\begin{figure*}[t]
    \centering
    \includegraphics[width=\textwidth]{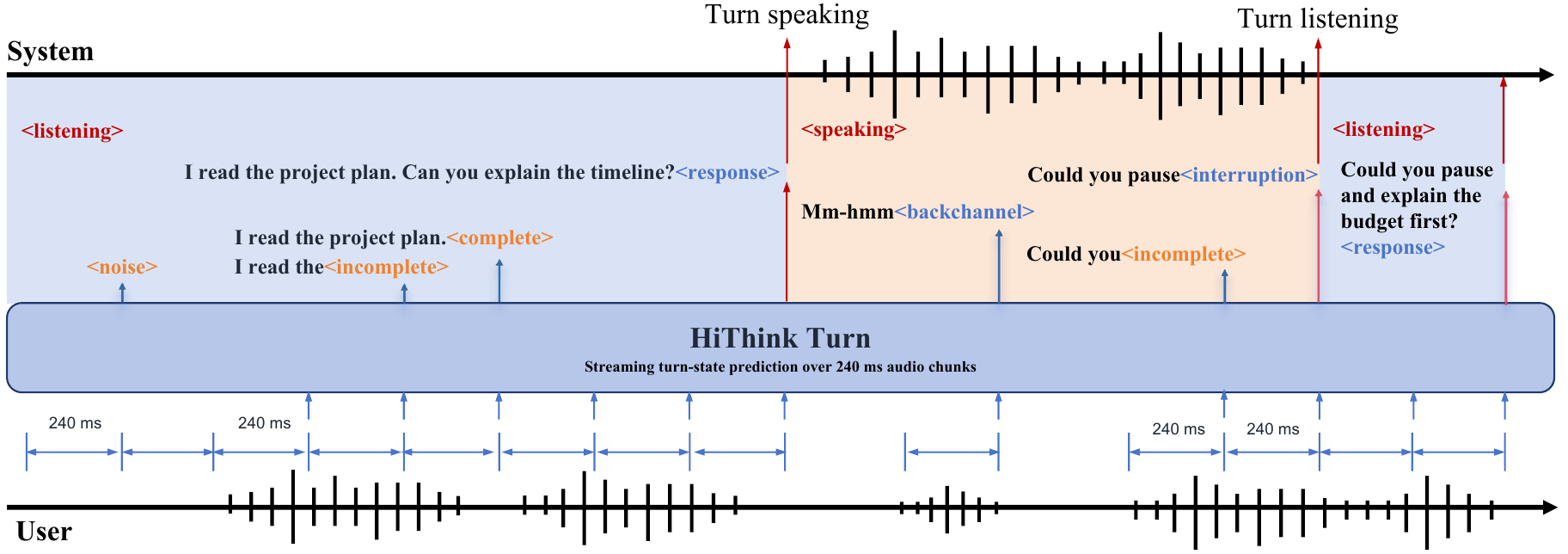}
    \vspace{-3pt}
    \caption{Overview of streaming turn-state inference in HiThink Turn, a module designed for cascaded full-duplex spoken dialogue systems that predicts turn states based on the system playback state and user interaction intent.}
    \label{fig:hithink_turn}
\end{figure*}

\section{HiThink Turn}
\label{sec:hithink_turn}

\subsection{Overview}

As shown in Fig.~\ref{fig:hithink_turn}, eight state labels are defined along three dimensions---audio validity, user interaction intent, and system playback state---to enable fine-grained duplex control. To provide reliable supervision for these states, the annotation pipeline in Fig.~\ref{fig:workflow} is developed to distinguish \textit{response} from \textit{interruption}, while boundary-aware samples are constructed to improve robustness to truncated streaming inputs. Under this formulation, turn-state prediction is conditioned on whether the system is speaking or listening, and the model is efficiently implemented by adapting Qwen3-ASR with LoRA.

\subsection{Intent-Aware Turn-State Design}

As summarized in Table~\ref{tab:state_labels}, turn states span three dimensions: audio validity, user intent, and system playback state. Following VAD and turn-detection models~\cite{pipecatai2025smartturn,liao2025flexduo}, the audio states identify extremely far-field or unintelligible input (\textit{noise}) and assess valid speech for semantic completeness (\textit{incomplete}/\textit{complete}). Since completeness alone cannot determine duplex actions, \textit{response}, \textit{backchannel}, and \textit{interruption} explicitly model user intent for finer-grained turn control.

Prediction is further conditioned on the system's \textit{speaking} or \textit{listening} state to account for playback-dependent user intent. In particular, \textit{backchannel} and \textit{interruption} are valid only during system playback.

\begin{table}[t]
\centering
\caption{Definitions of turn-state labels and playback states.}
\label{tab:state_labels}
\footnotesize
\setlength{\tabcolsep}{2.5pt}
\renewcommand{\arraystretch}{1.05}
\begin{tabularx}{\columnwidth}{
    @{}
    l
    l
    >{\raggedright\arraybackslash}X
    @{}
}
\toprule
\textbf{Dimension} & \textbf{Label} & \textbf{Definition} \\
\midrule

\multirow{3}{*}{Audio}
& \texttt{<noise>}
& Noise, invalid or unintelligible speech \\

& \texttt{<incomplete>}
& Valid speech with incomplete acoustic or semantic evidence \\

& \texttt{<complete>}
& Complete speech without response intent \\

\midrule

\multirow{3}{*}{User}
& \texttt{<response>}
& Complete speech requiring a system response or action \\

& \texttt{<backchannel>}
& Brief listener feedback that preserves system playback \\

& \texttt{<interruption>}
& Minimal intent-sufficient prefix warranting interruption \\

\midrule

\multirow{2}{*}{System}
& \texttt{<speaking>}
& System playback is in progress \\

& \texttt{<listening>}
& System is awaiting user input \\

\bottomrule
\end{tabularx}
\end{table}

\subsection{Turn-State Annotation Workflow}

\begin{figure}[!t]
    \centering
    \includegraphics[width=\columnwidth]{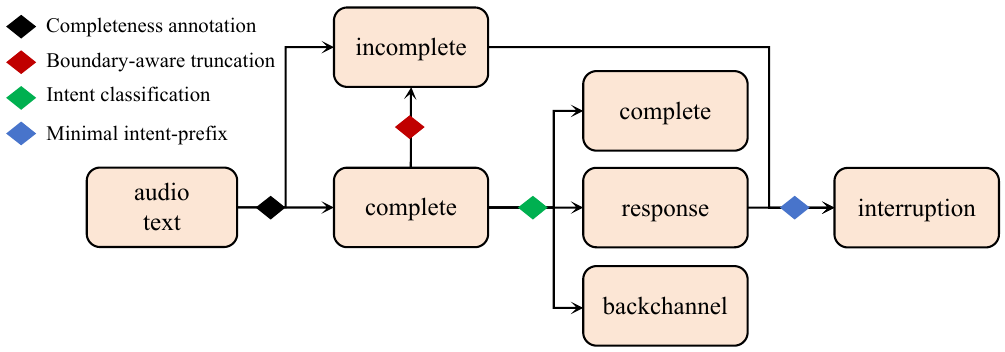}
    \makeatletter
    \patchcmd{\@makecaption}{\vskip 10pt}{\vskip 4pt}{}{}
    \makeatother
    \caption{Annotation Workflow for State Labels}
    \label{fig:workflow}
\end{figure}

Full-duplex data are scarce, and most public datasets are synthetic. Fig.~\ref{fig:workflow} illustrates the turn-state annotation pipeline used to construct the training data for HiThink Turn.

\noindent\textbf{Completeness annotation.}
EoT%
\footnote{\raggedright
\url{https://blog.speechmatics.com/semantic-turn-detection}\par}
assigns each transcript a completeness score by summing the
next-token probabilities of an end marker and sentence-final
punctuation. Samples scoring above 0.25 are labeled
\textit{complete}, while those below 0.01 are labeled
\textit{incomplete}. Cases in $[0.01, 0.25]$ are reviewed by
Qwen3.5-122B%
\footnote{\raggedright
\url{https://huggingface.co/Qwen/Qwen3.5-122B-A10B}\par}.

\noindent\textbf{Boundary-aware truncation.} To improve robustness to partial streaming audio, additional \textit{incomplete} samples are constructed at 240-ms chunk boundaries. For each audio--transcript pair $(\mathbf{x},\mathbf{y})$, forced alignment provides the end time $\tau_i$ of textual unit $y_i$. With $\Delta=0.24$~s, a streaming step $k$ is randomly selected before the final chunk. The corresponding incomplete pair is constructed as
\begin{equation}
\begin{gathered}
(\widetilde{\mathbf{x}}_k,\widetilde{\mathbf{y}}_k)
=
(\mathbf{x}_{[0,k\Delta]},\mathbf{y}_{1:m_k}),\\
m_k
=
\max\{i \mid \tau_i \leq k\Delta\}.
\end{gathered}
\label{eq:boundary_augmentation}
\end{equation}

Here, $m_k$ denotes the number of textual units completed before the
sampled streaming boundary. All resulting partial
audio--transcript pairs are assigned the \textit{incomplete} state.

\noindent\textbf{Intent classification.}
Semantically complete utterances may serve different interactional purposes. We therefore use an LLM to refine their labels from transcripts. Requests directed at the assistant for an answer or action are relabeled \textit{response}, while brief acknowledgments and listener feedback during system playback are labeled \textit{backchannel}. The remaining utterances are complete but do not solicit a response or warrant interrupting playback; they retain the \textit{complete} label.

\noindent\textbf{Minimal intent-sufficient prefix.} In barge-in scenarios, waiting for semantic completion delays
interruption. Inspired by simultaneous machine translation (SiMT)~\cite{fu2025east},
we identify the minimal intent-sufficient prefix classified as requiring a response, enabling earlier interruption without waiting
for utterance completion. This prefix is defined as
\begin{equation}
k
=
\min
\left\{
k \mid
f_{\mathrm{LLM}}\!\left(\mathbf{y}_{1:k}\right)
=
\mathrm{response}
\right\},
\label{eq:earliest_response}
\end{equation}
where $\mathbf{y}_{1:k}=(\mathbf{y}_1,\ldots,\mathbf{y}_k)$ denotes the accumulated prefix of complete semantic phrases obtained through EAST
\footnote{\url{https://github.com/biaofuxmu/EAST}}
segmentation, and $f_{\mathrm{LLM}}$ is the LLM-based response-intent classifier. The earliest prefix classified as expressing response intent, $\mathbf{y}_{1:k}$, is labeled \textit{interruption}, with its aligned end time marking the barge-in point. These partial pairs are initially labeled \textit{incomplete} and subsequently screened by an LLM, those expressing response intent undergo minimal intent-sufficient prefix annotation.

\noindent\textbf{Invalid speech and filtering.} 
 Unintelligible and far-field speech is collected and segmented into clips as \textit{noise} samples to improve noise rejection in realistic conversations. Annotated samples with identical transcripts but conflicting state labels are removed, and the number of samples sharing the same transcript and state label is capped to limit repetition and improve data balance.

\subsection{State-Conditioned Training}
\label{sec:state_conditioned_training}

During training, the system playback state is provided as conditioning input and excluded from loss computation. Since backchannels and interruptions occur only during system playback, \textit{backchannel} and \textit{interruption} samples are conditioned exclusively on \textit{speaking}, while all other samples are randomly conditioned on \textit{speaking} or \textit{listening}. The resulting prediction is formulated as
\begin{equation}
(\hat{\mathbf{y}},\hat{s})
=
\arg\max_{\mathbf{y},s}
p_{\theta}(\mathbf{y},s \mid \mathbf{x},s^{\mathrm{sys}}),
\label{eq:state_conditioned_prediction}
\end{equation}
where $\mathbf{x}$ denotes the input audio, $s^{\mathrm{sys}}$ is
the system playback state, and $\hat{\mathbf{y}}$ and $\hat{s}$
are the predicted transcript and turn state, respectively.

\begin{table}[t]
\centering
\makeatletter
\patchcmd{\@makecaption}{\vskip 10pt}{\vskip 4pt}{}{}
\makeatother
\caption{Full-clip accuracy (\%) on Easy Turn test sets.}
\label{tab:easy_turn}
\footnotesize
\setlength{\tabcolsep}{2.8pt}
\renewcommand{\arraystretch}{1.00}
\begin{tabular*}{\columnwidth}{
    @{\extracolsep{\fill}}
    c l c c c
    @{}
}
\toprule
\textbf{Lang.}
& \textbf{Model}
& \textbf{Complete}
& \textbf{Incomplete}
& \textbf{Avg.} \\
\midrule

\multirow{4}{*}{EN}
& SenseVoice + TEN
& \textbf{95.6}
& 76.6
& 86.1 \\
& SoulX-Duplug
& 77.7
& 89.0
& 83.4 \\
& X2-Turn
& 92.1
& 84.6
& 88.4 \\
\cmidrule(lr){2-5}
& \textbf{HiThink Turn}
& 87.7
& \textbf{96.0}
& \textbf{91.9} \\

\midrule

\multirow{4}{*}{ZH}
& Easy Turn
& 96.3
& 97.6
& 97.0 \\
& SoulX-Duplug
& 89.3
& 79.3
& 84.3 \\
& X2-Turn
& 91.0
& 93.0
& 92.0 \\
\cmidrule(lr){2-5}
& \textbf{HiThink Turn}
& \textbf{99.0}
& \textbf{99.7}
& \textbf{99.4} \\

\bottomrule
\end{tabular*}
\end{table}

\section{Experiments}
\label{sec:experiment}

\begin{table*}[t]
\centering
\makeatletter
\patchcmd{\@makecaption}{\vskip 10pt}{\vskip 4pt}{}{}
\makeatother
\caption{Results on Full-Duplex-Bench v1--v1.5. Rates are proportions
and latencies are in seconds. Best and second-best results are shown
in \textbf{bold} and \underline{underlined}, ablations shown in gray.}
\label{tab:fdb}
\small
\setlength{\tabcolsep}{1.0pt}
\setlength{\aboverulesep}{1pt}
\setlength{\belowrulesep}{1.5pt}
\renewcommand{\arraystretch}{0.90}

\begin{tabular*}{0.98\textwidth}{
    @{\extracolsep{\fill}}
    l
    *{9}{c}
    @{}
}
\toprule

\multirow{2}{*}{\textbf{Model}}
& \multicolumn{1}{c}{Pause Handling}
& \multicolumn{2}{c}{Turn Taking}
& \multicolumn{1}{c}{Backchannel}
& \multicolumn{2}{c}{Interruption v1}
& \multicolumn{2}{c}{Interruption v1.5}
& \multicolumn{1}{c}{Overall} \\

\cmidrule(lr){2-2}
\cmidrule(lr){3-4}
\cmidrule(lr){5-5}
\cmidrule(lr){6-7}
\cmidrule(lr){8-9}
\cmidrule(lr){10-10}

& \textbf{TOR}$\downarrow$
& \textbf{TOR}$\uparrow$
& \textbf{RL}$\downarrow$
& \textbf{RsR}$\uparrow$
& \textbf{TOR}$\uparrow$
& \textbf{RL}$\downarrow$
& \textbf{RpR}$\uparrow$
& \textbf{SL}$\downarrow$
& \textbf{Avg.}$\uparrow$ \\
\midrule

Moshi$^{\dagger}$
& 0.983
& \underline{0.941}
& \textbf{0.265}
& 0.060
& \textbf{1.000}
& \textbf{0.257}
& 0.500
& 1.160
& 0.504 \\

Freeze-Omni$^{\dagger}$
& 0.562
& 0.336
& 0.953
& 0.800
& 0.867
& 1.409
& 0.720
& 1.420
& 0.632 \\

Gemini Live$^{\dagger}$
& 0.283
& 0.655
& 1.301
& \underline{0.930}
& 0.891
& 1.183
& 0.330
& 2.200
& 0.705 \\

SoulX-Duplug
& 0.352
& 0.933
& \underline{0.511}
& 0.740
& \underline{0.970}
& 0.773
& 0.770
& \textbf{0.450}
& 0.812 \\

DuplexCascade
& \textbf{0.140}
& 0.832
& 1.724
& 0.782
& 0.955
& 1.225
& \underline{0.879}
& 1.503
& \underline{0.862} \\

X2-Turn
& \underline{0.224}
& 0.807
& 0.901
& 0.925
& 0.935
& \underline{0.659}
& 0.855
& \underline{0.574}
& 0.860 \\

\midrule

\textbf{HiThink Turn}
& 0.234
& \textbf{0.958}
& 0.625
& \textbf{0.959}
& \textbf{1.000}
& 1.177
& \textbf{0.980}
& 1.322
& \textbf{0.933} \\

\quad \textcolor{gray}{w/o boundary-aware data}
& \textcolor{gray}{0.262}
& \textcolor{gray}{0.842}
& \textcolor{gray}{0.605}
& \textcolor{gray}{0.935}
& \textcolor{gray}{0.990}
& \textcolor{gray}{1.142}
& \textcolor{gray}{0.980}
& \textcolor{gray}{1.293}
& \textcolor{gray}{0.897} \\

\quad \textcolor{gray}{w/o system state}
& \textcolor{gray}{0.336}
& \textcolor{gray}{0.782}
& \textcolor{gray}{0.636}
& \textcolor{gray}{1.000}
& \textcolor{gray}{1.000}
& \textcolor{gray}{1.324}
& \textcolor{gray}{0.970}
& \textcolor{gray}{1.410}
& \textcolor{gray}{0.883} \\

\bottomrule
\end{tabular*}

\vspace{1pt}
\begin{minipage}{0.98\textwidth}
{\small
\raggedright
TOR: take-over rate; RL: response latency; RsR: resume rate;
RpR: response rate; SL: stop latency.
Avg.: mean of five rate metrics, with Pause TOR replaced by
$1-\mathrm{TOR}$, latencies excluded.
$^{\dagger}$ Results from corresponding papers.
\par}
\end{minipage}
\vspace{-3pt}
\end{table*}

\subsection{Experiment Details}
\label{sec:experiment_details}

The training set comprises 7,633\,h of English and Chinese audio from ASR corpora, real-world QA recordings, and synthesized user turns from Duplex-Ultra\-Chat~\cite{zhang2024beyond}. Noise data come from WHAM~\cite{wichern2019wham}, CHiME-5~\cite{barker2018chime5}, and far-field recordings. The annotation pipeline in Fig.~\ref{fig:workflow} yields 5.552 million samples: 16.0\% \textit{noise}, 29.9\% \textit{incomplete}, 17.8\% \textit{complete}, 20.8\% \textit{response}, 14.8\% \textit{interruption}, and 0.7\% \textit{backchannel}. The noise subset contains 888k samples totaling 1,003\,h.

HiThink Turn is initialized from Qwen3-ASR-1.7B and adapted using LoRA
with rank $r=32$ through the ms-swift toolkit%
\footnote{\url{https://github.com/modelscope/ms-swift}}.
Training is conducted for two epochs on eight NVIDIA H100 GPUs using
AdamW with a learning rate of $1\times10^{-5}$ and a 5\% warmup ratio.
A per-device batch size of 32 and four-step gradient accumulation yield
an effective batch size of 1,024. The maximum sequence length is 4,096
tokens, and bfloat16 precision is used throughout training.

\begingroup
\setlength{\dbltextfloatsep}{6pt plus 1pt minus 1pt}
\subsection{Easy Turn Results}
We evaluate utterance-level completeness prediction on Easy Turn~\cite{li2026easy}. HiThink Turn achieves the highest macro accuracy in both English (91.9\%) and Chinese (99.4\%), together with the best accuracy on incomplete utterances. Misclassified complete English utterances often have little or no trailing silence. Appending 0.3\,s of silence corrects 61.5\% of these errors, suggesting that the model uses utterance-final acoustic cues in addition to transcript semantics.

\setlength{\parskip}{0pt}
\microtypesetup{activate=true}
\setlength{\textfloatsep}{4pt plus 1pt minus 1pt}
\setlength{\intextsep}{3pt plus 1pt minus 1pt}
\makeatletter
\patchcmd{\@makecaption}{\vskip 10pt}{\vskip 4pt}{}{}
\makeatother
\subsection{Full-Duplex-Bench Results}

Following SoulX-Duplug~\cite{soulxduplug2026}, we construct a cascaded full-duplex system and evaluate it on Full-Duplex-Bench v1--v1.5~\cite{lin2025fdbv1,lin2026fdbv1.5}. As shown in Table~\ref{tab:fdb}, HiThink Turn achieves the highest aggregate interaction score of 0.933 among the compared systems, with strong performance in turn taking, backchannel handling, and user interruption. With 240-ms chunks, mean model and end-to-end response latencies are 0.160\,s and 0.625\,s, respectively, remaining competitive among evaluated systems.
To evaluate minimal intent-sufficient prefixes, we hold the model and its predictions on Interruption v1.5 fixed and compare triggering on \textit{response} alone with triggering on either \textit{interruption} or \textit{response}. As shown in Fig.~\ref{fig:interruption_latency}, allowing \textit{interruption} triggers increases the response success rate from 89\% to 98\% and reduces mean decision latency among successful cases from 3.380\,s to 1.322\,s, a 60.9\% reduction. These results demonstrate that the additional trigger improves interruption success and timeliness without changing model predictions.

\begin{figure}[!htbp]
    \centering
    \includegraphics[width=\columnwidth]
    {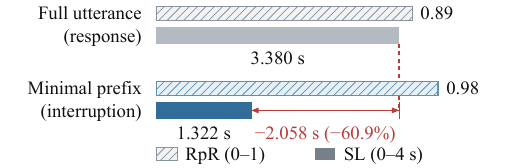}
    \caption{SL on Interruption v1.5. Minimal intent-sufficient prefix 
    decisions vs full-utterance decisions}
    \label{fig:interruption_latency}
\end{figure}

\subsection{Non-Target Speech Rejection Evaluation}

Non-target speech evaluation further assesses whether the system
maintains ongoing interaction when no response is warranted.
As shown in Table~\ref{tab:fdb_nontarget}, HiThink Turn achieves the
highest average RsR of 0.735 among the compared systems in FDB v1.5, demonstrating
its ability to suppress unnecessary interactions. It performs best on
Background but trails Gemini Live on Talk-to-Other, leaving room for
improvement in identifying system-directed speech with context.

\begin{table}[!htbp]
\centering
\caption{RsR for non-target-speech robustness on
Full-Duplex-Bench ($\uparrow$).}
\label{tab:fdb_nontarget}
\footnotesize
\setlength{\tabcolsep}{1.0pt}
\renewcommand{\arraystretch}{0.96}

\begin{tabular*}{\columnwidth}{
    @{\extracolsep{\fill}}
    l c c c
    @{}
}
\toprule
\textbf{Model}
& \textbf{Background}
& \textbf{Talk-to-Other}
& \textbf{Avg.} \\
\midrule

Moshi$^{\dagger}$
& 0.070
& 0.190
& 0.130 \\

Freeze-Omni$^{\dagger}$
& 0.250
& 0.250
& 0.250 \\

Gemini Live$^{\dagger}$
& \underline{0.300}
& \textbf{0.990}
& \underline{0.645} \\

\midrule

\textbf{HiThink Turn}
& \textbf{0.600}
& \underline{0.870}
& \textbf{0.735} \\

\bottomrule
\end{tabular*}
\end{table}

\subsection{Ablation Study}

We separately ablate boundary-aware data and system state context. Table~\ref{tab:fdb} shows that removing the former lowers the aggregate score despite slightly shorter latency. Error analysis links the latency reduction mainly to premature decisions when audio boundaries split words, leading to inappropriate waiting or early responses. This augmentation thus stabilizes predictions on partial speech.

Removing playback-state conditioning lowers turn-taking success, increases takeovers during pauses, and delays interruption, despite improving backchannel resume rate. The same request may require a response during listening but an interruption during playback. Since backchannel and interruption samples occur only under the speaking condition, removing this context merges speaking and listening samples, lowering the relative prevalence of interruption labels compared with the speaking condition and obscuring the distinction between responding and interrupting.

\section{Conclusion}
\label{sec:majhead}

We presented HiThink Turn, an intent-aware streaming turn-state predictor separating response intent from semantic completeness through playback context, minimal intent-sufficient prefix supervision, and boundary-aware training. Results on Full-Duplex-Bench and Easy Turn demonstrate strong turn control and completeness prediction, with interruption triggers reducing mean decision latency among successful cases by 60.9\% over response-only triggering. Future work will incorporate dialogue context and speaker cues to distinguish response intent, turn yielding, and system-directed speech.
\par
\clearpage
\endgroup

\bibliographystyle{IEEEbib}
\bibliography{refs}

\end{document}